\documentclass[letter,
               keeplastbox,   
               nospread,     
               ]{jacow}

\usepackage{pdfpages,multirow,ragged2e} 
\makeatletter%
	\ifboolexpr{bool{xetex}}
	 {\renewcommand{\Gin@extensions}{.pdf,%
	                    .png,.jpg,.bmp,.pict,.tif,.psd,.mac,.sga,.tga,.gif,%
	                    .eps,.ps,%
	                    }}{}
\makeatother

\ifboolexpr{bool{xetex} or bool{luatex}} 
 {}                                      
 {\usepackage[utf8]{inputenc}}           

\usepackage[USenglish]{babel}
\usepackage{physics}
\usepackage[utf8]{inputenc} 

\ifboolexpr{bool{jacowbiblatex}}%
 {%
  \addbibresource{jacow-test.bib}
  \addbibresource{biblatex-examples.bib}
 }{}
\begin{document}

\title{Predicting Beam Transmission Using 2-Dimensional Phase Space Projections Of Hadron Accelerators}

\author{Anthony Tran\thanks{tranant2@msu.edu}, Yue Hao, Michigan State University, East Lansing, USA \\
		 Brahim Mustapha, Jose L. Martinez Marin, Argonne National Laboratory, Argonne, IL, USA}
	
\maketitle

\begin{abstract}
We present a method to compressed the 2D transverse phase space projections from a hadron accelerator and use that information to predict the beam transmission. This method assumes that it is possible to obtain at least three projections of the 4D transverse phase space and that an accurate simulation model is available for the beamline. Using a simulated model we show that, a procedure using a convolutional autoencoder can be trained to reduce phase-space information which can later be used to predict the beam transmission. Finally, we argue that although using projections from a realistic non-linear distribution produces less accurate results, the method still generalizes well.
\end{abstract}

\section{Introduction}

A challenging problem in obtaining high beam power in hadron linacs, such as ATLAS, SNS, and FRIB, is understanding and minimizing uncontrolled beam loss, a major unexpected loss of the beam within the beamline.\cite{aleksandrov2016path}  In the low energy beam transport lines (LEBT), the beam must be carefully controlled to minimize the beam loss downstream. The beam is generally a collection of particles that can be described in six-dimensional space; three positions, and three momentum coordinates. For the DC beam in the LEBT, the longitudinal coordinates are not involve in the dynamics but just appear as parameters.  Therefore, each charged particle is described by its location in the four-dimensional (4D) transverse phase space ($x$, $x'$, $y$, $y'$), where primed coordinates are derivatives with respect to the longitudinal direction.

In the LEBT, multiple beam measurement devices such as Alison Scanners \cite{allison1983emittance}, Pepper-Pot emittance meters \cite{pikin2006pepper}, wire scanners \cite{koziol2001beam}, and viewers are used to capture one-dimensional(1-D) or two-dimensional (2D) profile measurements, which are projections of the four-dimensional (4D) transverse phase space.
Inferring the 4-D distribution from these projected 1-D and/or 2-D information is referred to as 4D tomography. Mathematical and physical methods, such as the maximum entropy principle \cite{reggiani2010transverse} \cite{wong20224d}, has been successfully demonstrated to realize the 4-D tomography in accelerators. However, there is still challenges in combining 1-D or 2-D information from different locations.  The optics deviation, which is when the machine deviates from the model, for example, will affect the accuracy of the 4-D tomography. 

In this paper, we tested a data-driven approach to predict the beam loss using 4D phase space distribution information 
encoded in a low dimensional vector from 2D projections measurements. The data was generated from virtual diagnostic instruments simulated using the beam dynamics code TRACK. The simulation is of a test lattice adopted from the LEBT of the ATLAS accelerator which consists of 6 quadruples and 5 virtual diagnostic instruments. The simulation results were used to develop a convolutional autoencoder to encode the data into a meaningful lower-dimensional representation, which will then be used to relate the phase-space information to the beam loss.

\section{Method}

\subsection{ATLAS Lattice}

\begin{figure}[ht]
    \centering
    \includegraphics[width=8cm]{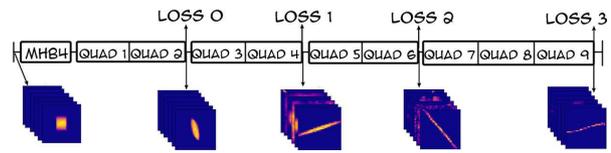}
    \caption{Cartoon of accelerator and beam measurements. The image shows where each beam measurement was collected from. }
    \label{fig:measurement}
\end{figure}

The presented study used data generated from the simulation of ATLAS's LEBT. The lattice used consists mainly of a multi-harmonic buncher, 9 quadruple magnets, 6 being used to tune the lattice, and 5 virtual diagnostic instruments. The virtual diagnostic instruments capture the 4D phase space of the beam, providing information on the 2D projections and beam transmission for use later. This amount of information is currently hard to achieve in a real accelerator but it is used to study the feasibility of the method, giving the initial model a higher chance to succeed.  

Figure \ref{fig:measurement} shows the location of the virtual diagnostic instruments. The measurements were recorded at five different locations: at the beginning, the end, and after quadrupoles 2, 4, and 6.

\subsection{Generating Data Using TRACK}
TRACK is a ray-tracing or particle-tracking code that can: (1) represent external fields accurately within the aperture. (2) calculate the particle coordinate at any point in the space. (3) determine calculate beam loss in both the ideal case and in the presence of complex field errors and device misalignments \cite{1591006}. 

\begin{table*}[h]
    \centering
    \caption{Parameter range used to generate data set of the initial beam distributions and quadruple settings.}
    \begin{tabular}{c|c}
    \hline
    Input\\
    \hline
    Voltages on Quadruples 1, 3, 5  &  uniform random number from [0,8] V\\
    Voltages on Quadruples 2, 4, 6   & uniform random number from [-8,0] V\\
    Initial Distribution & random distribution from 9 built in distribution\\
    $\epsilon_{x,y}$ & $0.12 + Normal(\mu=0, \sigma=0.012)$ cm*mrad \\
    $\alpha_{x,y}$ & $Normal(\mu=0, \sigma=1)$ unitless\\
    $\beta_{x,y}$ & $100 + Normal(\mu=0, \sigma=10)$ cm/rad\\
    \hline
    Output\\
    \hline
    Number of particles left  & [0,10000] particles. Taken at 4 different points\\
    Position of all particles  & Taken at 5 different points\\
    \hline
    \end{tabular}
    \label{tab:data}
\end{table*}

Since machine time is costly, the TRACK simulation was used to gather data. It was generated on Michigan State University's high-performance computing cluster for a week, producing over a million data points. As will be covered in a later section, a significant amount of data will be required for training autoencoders to high fidelity. The parameters for these simulations were varied according to Table \ref{tab:data}. These parameters were chosen within a small range so the model can interpolate well. Once the data was collected, it was filtered so that the initial beam distributions were contained within the beam aperture, resulting in a final data set with a size of around 430,000 simulation points.

To extract the 2D phase-space projections from the 4D phase space, all the particles were deposited onto an $n\cross n$ grid using pairs of the coordinates axes, $(x, x', y, y')$. This resulted in 6 independent projections.

\subsubsection{Non-linear Field}

Another data set was created to test the generalizability of the model which will be explained later. This was done with a perturbation to the initial distribution. The distribution was created simply by putting a non-linear magnetic field, such as a sextupole, at the beginning of the simulation. All other generative settings were kept the same. The could simulate the realistic case of ECR beams which experience sextupole field inside the source.

\subsection{Autoencoder}
An autoencoder is a nonlinear data reduction algorithm used in machine learning. It is composed of two parts, an encoder, and a decoder. The encoder takes a large input and reduces it into a lower dimension, known as a latent dimension, while the decoder attempts to reconstruct the latent dimension back into the original input. The error, which is the difference between the original and reconstructed data quantifies how well the latent dimension explains the original input. The advantage of compressing the data into a meaningful representation \cite{bank2020autoencoders} makes it more efficient to train a neural network model on the reduced data.

In the model, a convolutional autoencoder was implemented in PyTorch \cite{paszke2019pytorch} to reduce the dimension of the 2D projections of the phase space. A convolutional autoencoder uses a convolutional neural network as the encoder and decoder. A convolutional neural network is a type of neural network used to analyze visual information \cite{zhang2021dive,Scheinker_2021}. This has the advantage over principal component analysis\cite{abdi2010principal}, another data reduction algorithm, in that it includes spacial information, and can account for non-linear effects by using non-linear activation functions in the network. Activation functions take the output of a layer in a network and map it onto a set range. It was found that the reLu activation function and eLu activation function were the best activation functions to use \cite{ding2018activation}, which in this case, helps the model to train fast and be less likely to fail during training.

Each of the six 2D projections was given its own autoencoder. The decoder was able to reproduce all the original projections with reasonable accuracy, verifying that the projections were effectively encoded into a latent dimension. The latent dimensions sizes used for this paper were 32 for the $(x, x')$, and $(y, y')$ projection, and 16 for the rest. Given that the original images were made to be $33\cross33$, the inputs were significantly reduced.

\subsection{Modeling}

\begin{figure}[ht]
    \centering
    \includegraphics[width=8cm]{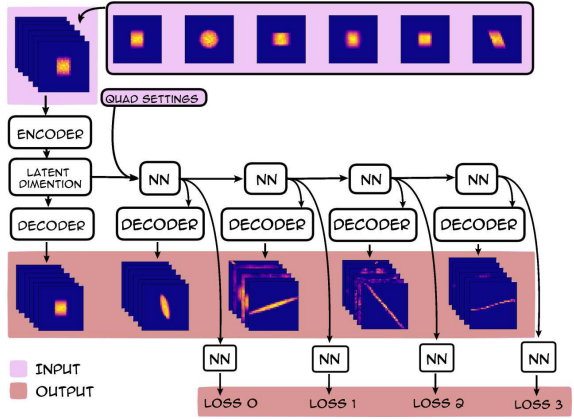}
    \caption{Cartoon of architecture. During training, the model takes all the 2D projections and loss value as input into the training. During testing, only the initial 2D projections were given and the model predicts the loss values and 2D projections in addition.}
    \label{fig:model}
\end{figure}

A neural network was used to create a surrogate model of the ATLAS front-end as shown in Figure \ref{fig:model}. The architecture is composed of first an encoder-decoder block to reduce each of the six phase-space projections into lower latent dimensions and then concatenated together. The quad settings were also concatenated onto this vector and then processed through fully connected layers. Each fully connected layer attempts to model the phase space changes by transforming the latent variables into a different set of latent variables which would describe the new 4D phase space. This new vector would be the input into the next fully connected layer to model the next transformation, a decoder block to reconstruct the distribution as 2D phase-space projections at that location, and another fully connected layer to predict the beam transmission represented by the number of particles left.

The encoder-decoder block uses a convolutional autoencoder as described in the previous section. The decoder was built similarly to the encoder, the only difference being that some adjustments were made to match the original dimensions. 
A decoder was not trained for every location, but it was combined for each projection. This saves limited GPU memory and produces a more generalized decoder.

To calculate the number of particles left, a two-layer fully connected network was used. Again, the network was not trained at every location, but it was combined to make a generalized particle loss predictor for the same reasons stated above. 

\subsection{Training}

Overall, the model encodes each initial phase-space projections into separate latent dimensions, then attempts to recreate the phase-space projections and predict the beam transmissions at the other 4 locations, and then finally compares them to the ground truth from the dataset. Using both results in a training loop, the model will update itself using gradient descent \cite{zhang2021dive} to better recreate the image in the next iteration. Gradient descent is performed using backpropagation on PyTorch computational map. The basic idea is that, using the gradient information, parameters in the network will be adjusted in the direction that will result in the final output being closer to the true values. The size of the step taken is related to the learning rate.

To quantify the difference between the predicted and ground truth results, a loss function was used. Commonly, mean square error loss is used because it punishes large deviations; however, this resulted in an overflow, an error that occurs when a computer produced a number larger than it can represent, in the gradient calculation during training. Absolute loss (L1 Loss) was used in those cases. This could also have been resolved by using a double or a long float, but this would use up valuable memory on the GPU.

To also aid in training, frozen layers were implemented to decrease training time and prevent gradient overflow. Frozen layers are layers where the gradient information was disconnected, thus preventing changes to that layer during training. This method was implemented in the following three-step training procedure:

\begin{enumerate}
    \item  The fully connected layers to train the loss predictor was frozen and trained with a learning rate of $0.01$. Thus only the encoder-decoder and the fully connected layers between decoders were trained.
    \item Frozen gradients were unfrozen, and unfrozen gradients were frozen, thus only the loss predictor was trained. The learning rate was also $0.01$
    \item Everything was unfrozen and trained at a lower learning rate of $0.001$.
\end{enumerate}

A problem that arises from using simulation is overfitting, which is a state where the model memorizes the training data rather than generalizing it. Since simulation generally is different from the actual machine due to installation and operation errors, if the model is not generalized or adjusted enough, it will perform poorly on the actual machine. 

This last step was used to better connect the two parts which also helps to prevent overfitting. A lower learning rate compared to the previous two steps is to discourage radical changes which may destroy the learned model in the first two steps.

\subsection{Results}

The model was tested on a newly generated dataset using the original parameters, as well as a non-linear dataset generated using a sextupole. Only the initial distributions were given, but the model would still predict the 2D projections and beam transmission at the other locations downstream. Then, to test the generalization of the model, a nonlinear field in the form of a sextupole \cite{lee2018accelerator} was added to the beginning of the simulation to generate a dissimilar subset of inputs.

For reference, an error of less than $1\%$ for results within 2 standard deviations from the mean would be sufficiently good for the prediction of the loss on ATLAS since it is a low power machine. For the rest of the paper, the percentage refers to a 2 standard deviation bound. The error is defined as the absolute difference between the ground truth and the predicted values divided by the total number of particles. The obtained values were plotted in Figure \ref{fig:results}A as a correlation graph. If there were no error, then there would be a perfectly straight line. Given that we have $10^4$ particles, the error for the original data set using six projections would be $3\%$.

This was then tested on the nonlinear sextupole distribution with fair results, an error of $5.5\%$ as shown in Figure \ref{fig:results}C. The model was able to generalize fairly well, however, it is still far from the ideal case. In this case, a machine learning model mainly interpolates the results, so the accuracy of a model depends on how much data it can train on. The more data points a model has, the better the interpolation will be.

Due to the nature of hadron accelerators, many of the quadruple configurations would produce a high particle loss because only a few configurations would allow most of the particles to pass. Thus, most of the dataset would be skewed towards high loss, resulting in higher accuracy in those cases since there is more data in those cases. To analyze this effect, the dataset was split into bins and as expected, the bin of particle loss between $9000-10000$ has an error around $2.5\%$ and for the bin of particle loss between $0-1000$, the error was as high as $5\%$. 

\subsubsection{Testing on a Smaller Data Set}

\begin{figure}[ht]
\begin{center}
\includegraphics[width=8cm]{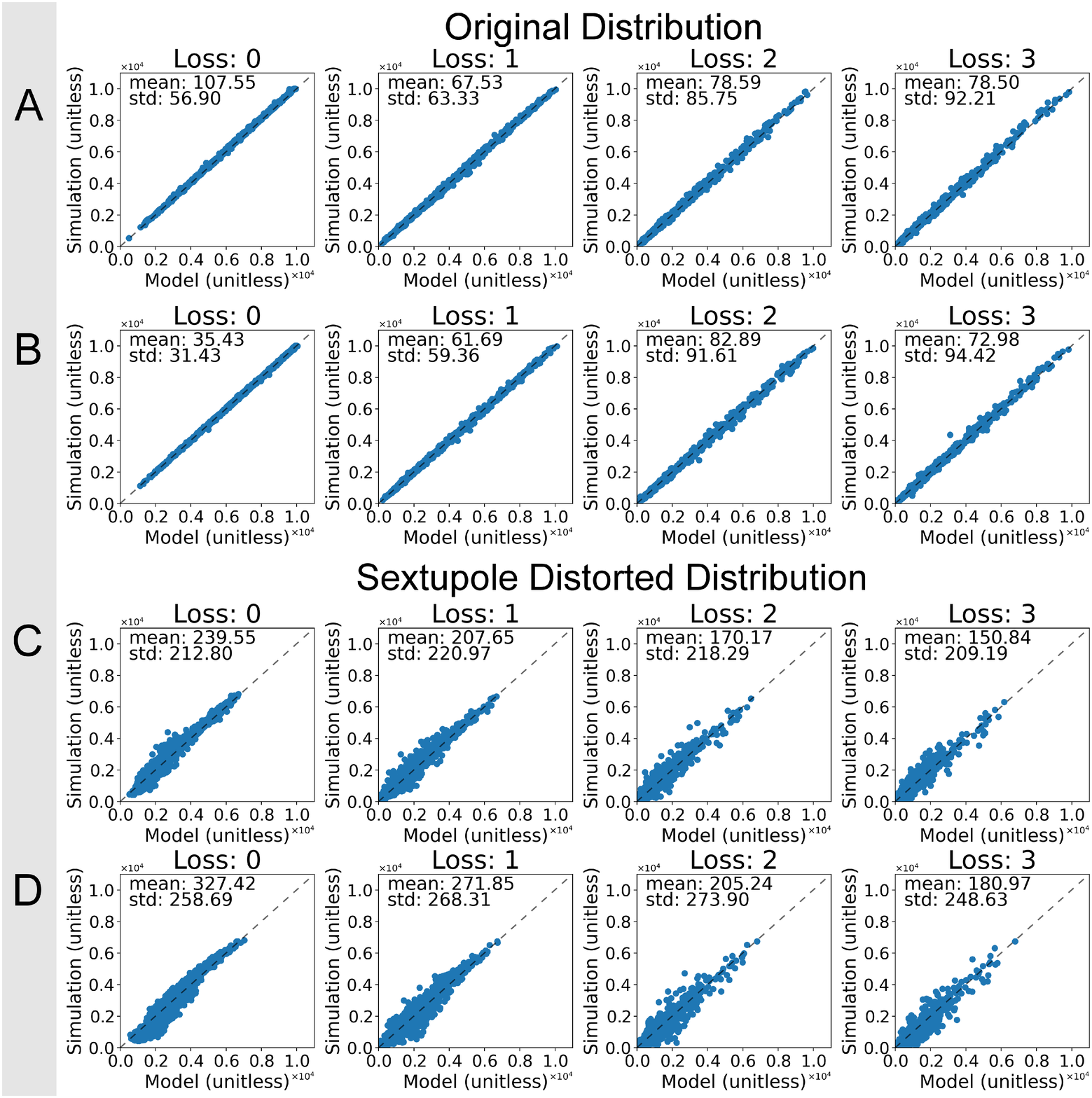}
\end{center}
\caption{\textbf{(A)} Histogram of original data set using six projections, \textbf{(B)} and same model but using three projections.\textbf{(C)} Histogram of original data set using six projections, \textbf{(D)} and same model but using three projections.}\label{fig:results}
\end{figure}

The same model was tested again, but with the $(x,y'), (x',y'),$ and $(y,x')$ projections removed. In Figure \ref{fig:results}B, the error predictions from the original data set show an improvement in the accuracy for "Loss: 0" while it has around the same error for the other losses. This is likely due to overfitting as the predictions from the non-linear data set show a loss of accuracy overall as seen in Figure \ref{fig:results}D; however, the model was shown to work with half the image data used, making this model more practical.

\section{Discussion}

A proof-of-principle machine learning based model has been reported to test a ML-based 4-D tomography using its 2-D projections and the capability to predict the beam transmission.  The result shows that if given only three projections of the 4D phase space, the projections can be reduced into a smaller latent dimension that contains the core information, which can then be used to predict the beam transmission downstream. If fewer projections were used, the model would not have enough information to describe the full 4D phase space. The latent dimension was verified to have contained the core information through a decoder that correctly reconstructed the encoded images. This method generalizes fairly well to initial beam distributions with non-linear perturbations, showing robustness and the potential for modeling the real machine.

Before bringing this to a real machine, it should be noted that this is a simplified model of an actual accelerator and there are multiple differences to consider. First, this model assumes that the accelerator elements can be modelled by a single parameter. Therefore, more complicated effects, such as misalignment of the magnets and the longitudinal overlapping of transverse magnets, are not considered.  Second, the model assumes the 2-d projections can be precisely measured and the measurement errors are not considered. The measurement error at various locations can be reduced by multiple measurements with different optics, which is not included in our simplified model. 

In this study, the beam loss value is represented by the percentage of the total beam intensity.  Accuracy of few percent beam loss has been demonstrated with this simplified model.  However, characterizing the beam loss with certain percentage of the total power results in a very different threshold for accelerators with different goal of beam power.  Therefore, the absolute power loss should be used in later applications for real accelerators. 

The model could be trained using data from a real machine, but collecting a large amount of data is expensive and time-consuming, thus models will have to be trained first on realistic simulation models and then transferred to the real machine. Methods known as “transfer learning” allow knowledge learned from the source dataset to be transferred to a target dataset \cite{zhang2021dive}. This could be done, for example, by freezing the model, adding another layer to the encoder and decoder, and training that layer on the distribution from the machine to adapt the model to the machine. Then, the rest of the model can be unfrozen and trained with a much smaller learning rate in order to fine-tune the model. Further studies on these will allow models to be trained first with simulations and then transferred to a machine.

Finally, this work assumes no accelerator knowledge, but further research will involve incorporating physics into the model. Some ways this could be done is by encoding constraints in the loss function during model training or by incorporating domain knowledge by including the transfer matrices in the calculation as a prior. The positive results of this work give hope that incorporating this knowledge may time time, increase sample efficiency, and further reduce the beam transmission error.

\section*{Conflict of Interest Statement}

The authors declare that the research was conducted in the absence of any commercial or financial relationships that could be construed as a potential conflict of interest.

\section*{Author Contributions}

AT did the data generation, the analysis, and the writing of this paper. YH advised AT along the way and revised the paper. BM helped revised the paper for better understanding. JM created the initial wrapper to run TRACK in python.

\section*{Funding}
This work was supported by the U.S. Department of Energy, under Contract No. DE-AC02-06CH11357. This research used the ATLAS facility, which is a DOE Office of Nuclear Physics User Facility.

\section*{Acknowledgments}
I would like to acknowledge Jaturong Kongmanee for multiple discussion related to machine learning.

\section*{Data Availability Statement}
The dataset for this study is available upon request.


\end{document}